\documentclass[prl,twocolumn,showpacs,superscriptaddress,10pt]{revtex4-1}
\usepackage[colorlinks,citecolor=blue,linkcolor=red,dvipdfm]{hyperref}
\usepackage{amsmath,amssymb,graphicx}
\usepackage{times}

\begin{document}
\title{Gap solitons in parity-time symmetric moir\'{e} optical lattices }

\author{Xiuye Liu}
\affiliation{State Key Laboratory of Transient Optics and Photonics, Xi'an
Institute of Optics and Precision Mechanics of CAS, Xi'an 710119, China}
\affiliation{University of Chinese Academy of Sciences, Beijing 100049, China}

\author{Jianhua Zeng}
\email{\underline{zengjh@opt.ac.cn}}
\affiliation{State Key Laboratory of Transient Optics and Photonics, Xi'an
Institute of Optics and Precision Mechanics of CAS, Xi'an 710119, China}
\affiliation{University of Chinese Academy of Sciences, Beijing 100049, China}


\begin{abstract}
Parity-time ($\mathcal{PT}$) symmetric lattices have been widely studied in controlling the flow of waves, and recently moir\'{e} superlattices, connecting the periodic and non-periodic potentials, are introduced for exploring unconventional physical properties in physics; while the combination of both and nonlinear waves therein remains unclear. Here, we report a theoretical survey of nonlinear wave localizations in $\mathcal{PT}$ symmetric moir\'{e} optical lattices, with the aim of revealing localized gap modes of different types and their stabilization mechanism. We uncover the formation, properties, and dynamics of fundamental and higher-order gap solitons as well as vortical ones with topological charge, all residing in the finite band gaps of the underlying linear-Bloch wave spectrum. The stability regions of the localized gap modes are inspected in two numerical ways: linear-stability analysis and direct perturbed simulations. Our results provide an insightful understanding of solitons physics in combined versatile platforms of $\mathcal{PT}$ symmetric systems and moir\'{e} patterns.
\end{abstract}

\maketitle

\section{Introduction}

MMoir\'{e} pattern---a periodic pattern overlaps its copy with a relative twist---as a novel two-dimensional (2D) material have shown great unique physical properties in condensed matter physics, including tunable flat bands, superconductivity and correlated insulator phases at twist (magic) angles in twisted double-layer graphene~\cite{TBG1,TBG2,TBG3,TBG4}, leading to an emerging field called twistronics---manipulating the electronic properties through the relative twist angle~\cite{Twistronics}. Recently,  the studies of moir\'{e} patterns and physics have entered regimes of optics and photonics: particularly, reconfigurable photonic moir\'{e} lattices were created in 2D photorefractive media by optical induction~\cite{PL-induced}, and localization-delocalization transition of light~\cite{opt1} and  optical solitons formation induced by twisting angle~\cite{opt2,SP16} were observed respectively in linear and nonlinear contexts; magic-angle lasers with unique confinement mechanism were fabricated in nanostructured moir\'{e} superlattices~\cite{opt-laser}; multifrequency solitons generation in quadratic nonlinear media with commensurate-incommensurate photonic moir\'{e} lattices was predicted~\cite{Quadratic}; to name just a couple of examples.

In addition, moir\'{e} optical lattices have also been proposed in the contexts of ultracold atoms and the associated moir\'{e} physics therein are being revealed~\cite{BEC-moire1,BEC-moire2,BEC-moire3}: to be specific, simulating twisted bilayers is possible by using cold atoms in state-dependent optical lattices, that show Dirac-like physics and band narrowing feature, enabling them an ideal candidate to observe similar physics (like strongly correlated phenomena in condensed matters) with larger rotation angles~\cite{BEC-moire1}; simulating twistronics without a twist---a highly tunable scheme, which rules out a physical bilayer or twist, to synthetically emulate twisted bilayer systems in the setting of ultracold atoms trapped in an optical lattice was proposed~\cite{BEC-moire2};  moir\'{e} physics including tunable flat bands and Larkin-Ovchinnikov superfluids in spin-twisted optical lattices (rather than bilayers) was investigated~\cite{BEC-moire3}; and considering the fact that electromagnetically induced regular optical (or photonic) lattices via atomic coherence in atomic ensembles are mature in experiments, we recently proposed a related scheme to create electromagnetically induced moir\'{e} optical lattices in a three-level coherent atomic gas (either hot or cold atoms) in the regime of electromagnetically induced transparency~\cite{FOP}. Consequently, we can safely conclude that optics and ultracold atoms open flexible and promising routes toward the realization of moir\'{e} optical lattices and associated moir\'{e} physics.

Moir\'{e} patterns bridge the gap between periodic structures and aperiodic ones, offering a new platform for studying nonlinear localization of light~\cite{opt2,SP16}. Particularly, the conventional periodic structures like photonic crystals and lattices in optics~\cite{PCF,PC,rev-light} and optical lattices in the context of ultracold atoms~\cite{OL-RMP,NL-RMP,rev-trapping} exhibit finite photonic or atomic band gaps~\cite{ABG}, the precise control of which and the corresponding nonlinearity could result into the emergence of a new spatially self-localized state called gap solitons (GSs) under repulsive (defocusing) nonlinearity
~\cite{GS1,GS4,GS5,GS6,GS8,GS9,BEC-darkGS,NL-focus,DarkGS-Q,DarkGS-CQ}. Experimentally, optical GSs have been confirmed in optical Bragg gratings~\cite{GS-FBG} and nonlinear photonic crystals~\cite{GS-WA}, and the creation of atomic GSs of Bose-Einstein condensates in optical lattices~\cite{GS-BEC}. It is worth noting that parity-time ($\mathcal{PT}$) symmetric lattices as an interesting periodic structure were heavily studied in optical~\cite{Bender-PT,Bender-Review,NPhoton-PT,NSR-PT,NM-PT,AM-PT} and matter-wave~\cite{PT-OL-3level,PT-atomOL-exp1} media and beyond in past years, providing a fertile land for investigating nonlinear waves including GSs~\cite{Soliton-PT,PRE-PT,RMP-PT,LPR-PT,Isci}. However, to the best of our knowledge, the combination of moir\'{e} patterns and  $\mathcal{PT}$ symmetry has not yet been reported.

In this work, we address the formation, properties, and dynamical stability of matter-wave GSs in a Bose-Einstein condensate (BEC) trapped with $\mathcal{PT}$ symmetric moir\'{e} optical lattices constituted of two 2D $\mathcal{PT}$ lattices with a twist (rotation) angle. Spatially localized nonlinear excitations of three kinds of coherent matter waves---fundamental GSs and high-order ones grouped as two fundamental modes, as well as gap vortices with topological charge---situating inside the first and second atomic band gaps of the linear Bloch-wave spectrum are found, underlining the tunable flat bands of the twisting $\mathcal{PT}$ periodic structure and robust stability of the reconfigurable nonlinear localized matter-wave structures within the associated gaps. The stability and instability properties of the nonlinear localized modes are assessed by linear-stability analysis and direct perturbed simulations, and both show a well agreement.


\section{Results and discussion}\label{sec2}

\subsection{Theoretical model}
\subsubsection{Gross-Pitaevskii equation}

The dynamics of a BEC cloud in 2D $\mathcal{PT}$ symmetric moir\'{e} optical lattices is described by the Gross-Pitaevskii (GP) equation for the dimensionless macroscopic wave function $\Psi$
\begin{equation}
i\frac{{\partial \Psi}}{{\partial t}} =  - \frac{1}{2}\nabla ^2 \Psi+ V_{\mathcal{PT} } ({\bf{r}})\Psi + \left| \Psi \right|^2 \Psi,
\label{GP}
\end{equation}
here Laplacian $\nabla ^2=\partial^2/\partial x^2+\partial^2/\partial y^2$ and $\textbf{r}=(x,y)$, the last term represents the nonlinearity where the repulsive atom-atom interactions controlled by Feshbach resonance  is chose. We stress that, in the context of nonlinear optics, the corresponding nonlinear Schr\"{o}dinger equation for the propagation dynamics of field amplitude is deduced readily by substituting the time $t$ with propagation distance $z$ in Eq.~(\ref{GP}). The $\mathcal{PT}$ moir\'{e} lattice under study yields
\begin{equation}
\begin{aligned}
V_{\mathcal{PT} } ({\bf{r}})=V_1[(\mathrm{cos}^2x+\mathrm{cos}^2y)+iV_0(\mathrm{sin}2x+\mathrm{sin}^2y)] \\
                                                     +V_2[(\mathrm{cos}^2x'+\mathrm{cos}^2y')+iV_0(\mathrm{sin}2x'+\mathrm{sin}2y')],
\end{aligned}
\label{OL}
\end{equation}
where $V_{1,2}$ being the strength and $V_0$ is the imaginary potential strength, and the strength contrast between the two sublattices is defined as $p=V_{2}/V_{1}$, setting $V_{1}=4$, for discussion. Note that the potential Eq.~(\ref{OL}) matches $\mathcal{PT}$ symmetry, $V_{\mathcal{PT}} ({\bf{r}})=V^\ast_{\mathcal{PT}} ({\bf{-r}})$, and it reduces to the usual (non-$\mathcal{PT}$) moir\'{e} lattice at $V_0=0$, and to the conventional $\mathcal{PT}$ lattice at $V_2=0$. The $(x, y)$ plane is related to the rotation $(x', y')$ plane with a twisting angle $\theta$
\begin{equation}\begin{split}
 \binom{x'}{y'} =\binom{\cos(\theta) {\kern 1pt}{\kern 1pt}-\sin(\theta)}{\sin(\theta) {\kern 1pt}{\kern 1pt}{\kern 1pt}{\kern 1pt}{\kern 1pt}{\kern 1pt}\cos(\theta)}\binom{x}{y} .
\end{split}\label{Rotation}\end{equation}
%
\begin{figure}[h]
\begin{center}
\includegraphics[width=0.8\columnwidth]{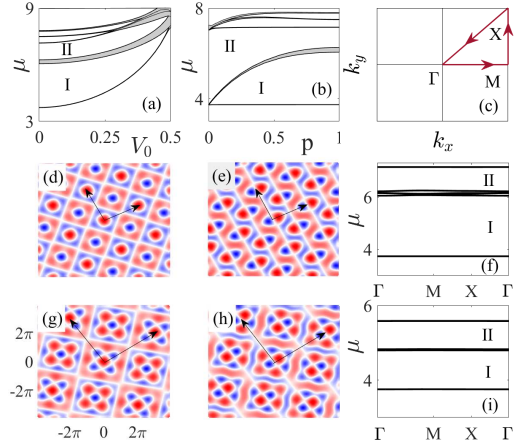}
\end{center}
\caption{Band-gap structures for the 2D $\mathcal{PT}$ symmetric moir\'{e} optical lattices at $\theta=\arctan(3/4)$ with increasing imaginary potential strength $V_0$ (a) and strength contrast $p$ (b). (c) The first Brillouin zone in the 2D reciprocal space. Contour plots of the real (d, g) and imaginary (e, h) parts of the lattice (shaded blue, lattice potential minima; shaded red, lattice potential maxima) at $\theta=\arctan(3/4)$ (d, e) and $\theta=\arctan(5/12)$ (g, h), their corresponding band-gap diagrams (f, i) at $V_0=0.02$ and $p=1$ in the reduced zone representation.  I and II in  (f, i) represent the first and second band gaps.}
\label{fig1}
\end{figure}
%

\subsubsection{Linear tunable flat-band properties}

Depicted in Fig.~\ref{fig1} is the linear Bloch spectrum of the $\mathcal{PT}$ moir\'{e} optical lattice [Eq.~(\ref{OL})] under Pythagorean angle $\theta=\arctan[2\alpha\beta/(\alpha^2-\beta^2)]$, with the Pythagorean triples $(\alpha^2-\beta^2, 2\alpha\beta, \alpha^2+\beta^2)$ for natural numbers $(\alpha, \beta)$. One can see from the Fig.~\ref{fig1}(a) that the widths of the first and second finite gaps compress rapidly with increasing $V_0$ at a defined angle $\theta$ ( i.e., $\theta=\arctan(3/4)$), and non-Hermitian degeneracy arises at an exceptional point (singularity) $V_0=0.5$, at which there is no any Bloch gap. With an increase of strength contrast $p$, more and more flat bands appear, making the widening of the first gap while the splitting of the second gap, according to Fig.~\ref{fig1}(b). For the 2D square $\mathcal{PT}$ moir\'{e} lattice at Pythagorean angle, the associated first reduced Brillouin zone in the reciprocal space is given in Fig.~\ref{fig1}(c). Typical real and imaginary parts of such lattices are shown in Figs.~\ref{fig1}(d) and~\ref{fig1}(e) at $\theta=\arctan(3/4)$, and in Figs.~\ref{fig1}(g) and~\ref{fig1}(h) at $\theta=\arctan(5/12)$, the corresponding linear band-gap structures are respectively shown in Figs.~\ref{fig1}(f) and~\ref{fig1}(i), where exist a broad first and second gaps.\vspace{5mm}

The stationary solution $\phi$ at chemical potential $\mu$ of the Eq.~(\ref{GP}) is given by $\phi=\Psi e^{-i\mu t}$, yielding
\begin{equation}
\mu \phi=-\frac{1}{2}\nabla ^2\phi+V_{\mathcal{PT} } ({\bf{r}})\phi+\left|\phi \right|^2\phi.
\label{station}
\end{equation}

Since we are interested in the GS solutions supported by the 2D $\mathcal{PT}$ moir\'{e} lattices, then their stability property is a key issue, which is evaluated by linear-stability analysis. Thus, we perturb the solution as $\psi=[\Phi+\rho\exp{(\lambda t)}+\varrho^{\ast}\exp{(\lambda^{\ast} t)}]\exp{(-i\mu t)}$, here $\Phi$ the unperturbed solution constructed from Eq.~(\ref{station}), $\rho$ and $\varrho$ being small perturbations under eigenvalue $\lambda$.  Substituting it into the Eq.~(\ref{GP}) would lead to the linear eigenvalue problem:
\begin{equation}\label{eig}
\left(
\begin{array}{cccc}
 L & \phi^2\\
 -\phi^{\ast 2} & -L
\end{array}
\right )
\left(
\begin{array}{cccc}
 \rho\\
 \varrho
\end{array}
\right )
=i\lambda
\left(
\begin{array}{cccc}
 \rho\\
 \varrho
\end{array}
\right ),
\end{equation}
with $L=-\mu-\frac{1}{2}\nabla^2+2|\phi|^2+V_{\mathrm{PT}}(\bf{r})$. One can see from Eq.~(\ref{eig}) that the solution is stable only when all the real parts of the eigenvalues are zero [$Re(\lambda)=0$]; it is unstable otherwise.\vspace{5mm}

\begin{figure}[h]
\begin{center}
\includegraphics[width=0.8\columnwidth]{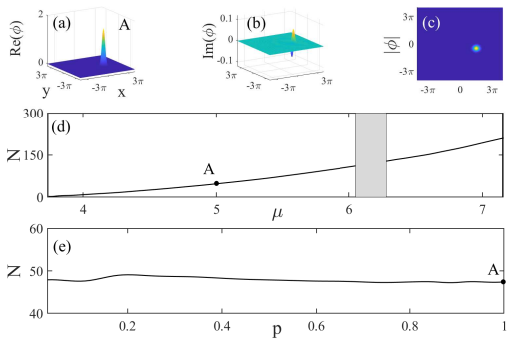}
\end{center}
\caption{Typical profile of a fundamental GS supported by the 2D $\mathcal{PT}$ symmetric moir\'{e} optical lattice at $\theta=\arctan(3/4)$ (a, b, c). The corresponding real (a) and imaginary (b) part, and contour plot of the module (c). Condensate population, $N$, as a function of chemical potential $\mu$ (d) and strength contrast $p$ (e) at $\theta=\arctan(3/4)$. Other parameters: $\mu=5, N=47.4 $ in (a, b, c).  }
\label{fig2}
\end{figure}

\subsection{Nonlinear Localized modes and their properties}
\subsubsection{Fundamental gap solitons}

The typical nonlinear localized mode in $\mathcal{PT}$ moir\'{e} optical lattices is the fundamental mode, matter-wave GSs, populated within the atomic finite gaps, characteristic profile of which is displayed in Figs.~\ref{fig2}(a),~\ref{fig2}(b), and~\ref{fig2}(c). It is observed that the real wavefunction,Re($\phi$), resembles a bright gap soliton, while its imaginary part, Im($\phi$), takes the form of dipole one. The condensate population, that is the number of atoms, $N{\rm{ = }}\int_{ -\infty}^{{+\infty}} {\left| \phi(\bf{r})  \right|} ^2 d{\bf{r}}$, as a function of the chemical potential $\mu$ for the GSs in the $\mathcal{PT}$ moir\'{e} lattices at $\theta=\arctan(3/4)$, is collected in Fig.~\ref{fig2}(d), displaying a `anti-Vakhitov--Kolokolov' (anti-VK) criterion, $dN/d\mu>0$, a necessary but not a sufficient condition for the stability of GSs in periodic structures with repulsive (defocusing) nonlinearity~\cite{BEC-darkGS,NL-focus,DarkGS-Q,DarkGS-CQ}. For a given $\mu$ inside the first gap, the dependency between number of atoms $N$ and strength contrast $p$, $N(p)$, is obtained in Fig.~\ref{fig2}(e), revealing a slight vibration of $N$ when altering $p$, such feature provides a flexible opportunity to launch fundamental GSs in moir\'{e} optical lattices with changeable strength contrast, in addition to the twisting angle $\theta$. A fact should be emphasized is that the fundamental GSs are very stable in the midst of the first and second gaps, they are unstable as long as they are excited near the both edges of the Bloch bands, and the direct perturbed evolutions shown in the following will prove it.\vspace{5mm}

\begin{figure}[h]
\begin{center}
\includegraphics[width=0.8\columnwidth]{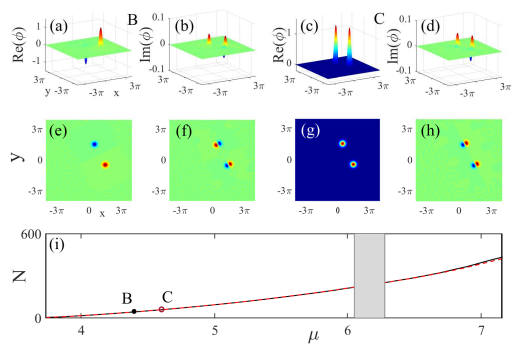}
\end{center}
\caption{Typical profiles of higher-order GSs grouped as two out-of-phase (a, b) and in-phase (c, d) fundamental GSs at $\theta=\arctan(3/4)$, the corresponding contour plots are displayed in second line (e$\sim$h) . (i) Condensate population, $N$, as a function of chemical potential $\mu$ at $\theta=\arctan(3/4)$ (Black, out-of-phase mode; red dashed, in-phase mode). Other parameters: $\mu=4.4, N=43.9 $ for B and $\mu=4.6, N=59.3 $ for C. }
\label{fig3}
\end{figure}

\subsubsection{Higher-order gap solitons}

Besides the fundamental mode reported in Fig.~\ref{fig2}, the $\mathcal{PT}$ moir\'{e} optical lattices can also support higher-order spatially localized gap modes that may be considered as composite structures of several fundamental GSs. Two examples of such higher-order modes, composing of two out-of-phase and in-phase fundamental GSs, are displayed in Figs.~\ref{fig3}(a, b) and Figs.~\ref{fig3}(c, d), respectively. Their contour plots are shown in the second line of Fig.~\ref{fig3}, with an emphasis on their tilted placements structured by the twisting optical lattices; and conform to the fundamental counterpart in Fig.~\ref{fig2}(b), the imaginary sections of the composite GSs emerge always as a dipole mode. The depending relations  $N(\mu)$ of both higher-order GSs modes are summed up in Fig.~\ref{fig3}(i), demonstrating, once again, the empirical stability anti-VK criterion, $dN/d\mu>0$~\cite{BEC-darkGS,NL-focus,DarkGS-Q,DarkGS-CQ}. \vspace{5mm}

\subsubsection{Gap vortices}

\begin{figure}[h]
\begin{center}
\includegraphics[width=0.8\columnwidth]{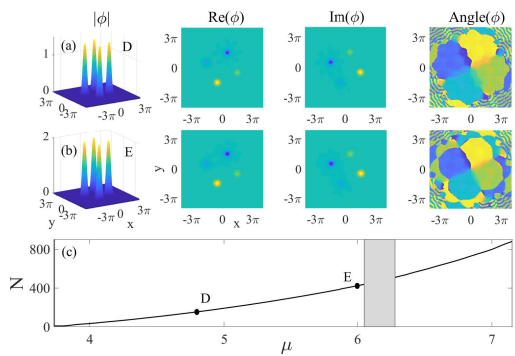}
\end{center}
\caption{Profiles of gap vortices consisted of four fundamental GSs with vortex charge $S=1$ prepared within (a) and near the upper edge (b) of the first finite gap, the corresponding condensate population, $N$, as a function of chemical potential $\mu$ at $\theta=\arctan(3/4)$ (c). The panels for the top and centre lines denote respectively, contour plot of the module, real and imaginary parts, as well as the associated phase structure. Other parameters for gap vortices marked by points D and E: (a) $\mu=4.8, N=153.5$; (b) $\mu=6, N=430$.}
\label{fig4}
\end{figure}

It is instructive to see whether the $\mathcal{PT}$ moir\'{e} optical lattices could sustain robust stable complex localized gap modes representing as gap vortices with topological charge (winding number) $S$. Our numerical calculations demonstrate that it is possible to create stable gap vortices with $S=1$ in such a novel periodic structure, provided that they are excited within the finite gaps. Typical profiles of the 2D hollow gap vortices with $S=1$ are in the form of four fundamental GSs entangled with $2\pi$ phase, according to the Figs.~\ref{fig4}(a) and~\ref{fig4}(b), where the corresponding real and imaginary parts of the wave functions and the associated phase structures are included. Evidently, both the real and imaginary parts show the similarity of having positive and negative values; and counterintuitively, the imaginary wave function Im$(\phi)$ for the vortex gap mode does not exhibit a dipole-like feature for each fundamental GSs, in contrary to their fundamental counterparts and higher-order ones as depicted respectively in Fig.~\ref{fig2}(b) and Figs.~\ref{fig3}(b, d, f, h). The non-dipole feature of imaginary part of the gap vortices may be explained by the unique and inherent property of localized vortical modes---the phase factor, which, together with the structural property of $\mathcal{PT}$ moir\'{e} optical lattices, determines the wave structures (both real and imaginary parts) of gap vortices. This can be clearly observed from the corresponding phase structures of the gap vortices displayed in the right columns of the first two lines of Fig.~\ref{fig4}. Also, in the Fig.~\ref{fig4}(c), we have obtained the curve $N(\mu)$ for such kind of gap vortices with topological charge $S=1$, showing an increase relationship.\vspace{5mm}

\subsubsection{Dynamics of gap solitons and vortices}

\begin{figure}[h]
\begin{center}
\includegraphics[width=0.9\columnwidth]{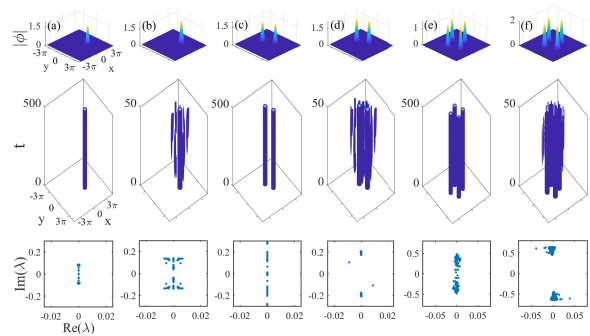}
\end{center}
\caption{Profiles of fundamental GSs (a, b), higher-order GSs (c, d),  and gap vortices with $S=1$ (e, f) prepared within the first (a, c, e, f) and second (b, d) finite gaps. The corresponding perturbed evolutions and linear eigenvalue spectra obtained from linear-stability analysis are displayed in the second and third lines, respectively. Other parameters: (a) $\mu=5, N=47.4$; (b) $\mu=6.29, N=128.2$; (c) $\mu=5.2, N=114.7$; (d) $\mu=6.28, N=252.6$; (e) $\mu=4.8, N=153.5$; (f) $\mu=6, N=430$.}
\label{fig5}
\end{figure}

The stability properties of all the 2D localized gap modes (fundamental and higher-order GSs, gap vortices) have been measured in linear-stability analysis and direct numerical simulations of the perturbed evolutions, using Eqs.~(\ref{eig}) and~(\ref{GP}) respectively, as pointed out previously. Our findings suggest the common feature of spatially localized modes supported by $\mathcal{PT}$ moir\'{e} periodic structures, that is, the GSs are stable within the middle portions of finite gaps, and are unstable close to the band edges. For the fundamental GSs in Figs.~\ref{fig5}(a, b), higher-order modes in Figs.~\ref{fig5}(c, d), and gap vortices at topological charge $S=1$ in Figs.~\ref{fig5}(e, f), with the stable ones being prepared in gaps and unstable ones near the band edges, their direct perturbed evolutions are depicted in the second line of Fig.~\ref{fig5}, which are in good agreement with their linear eigenvalue spectra produced by linear-stability analysis in the bottom line of Fig.~\ref{fig5}. It is important to emphasize that, during the evolution, the unstable localized gap modes develop multiple side peaks which reduce the necessary number of atoms ($N$) for sustaining the original gap modes; by contrast, shapes (good coherence) of stable modes can always keep.

\subsection{Experimental consideration}

Although our focus is merely on BECs loaded onto $\mathcal{PT}$ moir\'{e} optical lattices, the physical system under study could directly apply to the nonlinear optics context for describing light propagation in $\mathcal{PT}$ moir\'{e} photonic crystals and lattices, since both contexts share the same model~\cite{NLS-GP}; the only difference is to replace the time $t$ by propagation direction $z$ in Eq.~(\ref{GP}), and the chemical potential $\mu$ by propagation constant $-b$ in Eq.~(\ref{station}), then the wave function $\phi (x, t)$ is replaced by the electromagnetic field amplitude of laser pulse $E(z, x)$, and the number of atoms $N$ would be the soliton power $U$. Considering the fact that the $\mathcal{PT}$ optical lattices have been successfully fabricated in optics (and photonics)~\cite{Bender-PT,Bender-Review,NPhoton-PT,NSR-PT,NM-PT,AM-PT} and ultracold atoms
~\cite{PT-OL-3level,PT-atomOL-exp1}, the $\mathcal{PT}$ optical lattices of moir\'{e} type could also be easily realized using the current-state-of-the-art experimental technologies. And particularly, the GSs have been created in both contexts~\cite{GS-FBG,GS-WA,GS-BEC}, the spatially localized gap modes predicted here can thus be envisioned in both experimental platforms.\vspace{5mm}

\subsection{The materials and methods}
All the numerical results presented above obey the following numerical procedure: the stationary GS solution is firstly found from Eq.~(\ref{station}) via modified squared-operator iteration method~\cite{MSOM},  then its stability is measured by means of the linear-stability analysis [Eq.~(\ref{eig})] in Fourier collocation method~\cite{MSOM}, and direct perturbed simulations [Eq.~(\ref{GP})] using fourth-order Runge-Kutta method in real time.

\section{Conclusion}\label{sec3}

Concluding, we have addressed an as yet unresolved issue of the excitations and stability of 2D GSs supported by $\mathcal{PT}$ symmetric moir\'{e} optical lattices that exhibit tunable flat-band feature. Three categories of GSs are found, which are fundamental GSs and higher-order ones as well as gap vortices with topological charge, residing in both the first and second finite band gaps of the associated diffraction diagram, and their stability are evaluated in linear-stability analysis and direct perturbed simulations. The experimental platforms for observing them are discussed, and we envision that the localized gap modes in $\mathcal{PT}$ symmetric moir\'{e}  periodic structures are within the reach in contexts of nonlinear optics and atomic media.

\section*{Acknowledgements}

This work was supported by the National Natural Science Foundation of China (NSFC) (Nos. 61690224, 61690222,12074423); Major Science and Technology Infrastructure Pre-research Program of the CAS (No. J20-021-III); Natural Science Basic Research Program of Shaanxi  (No. 2019JCW-03); Key Deployment Research Program of XIOPM (No.S19-020-III).
~\\

\medskip
\textbf{Conflict of Interest} \par
The authors declare no conflicts of interest.


\begin{thebibliography}{0}%
\makeatletter
\providecommand \@ifxundefined [1]{%
 \@ifx{#1\undefined}
}%
\providecommand \@ifnum [1]{%
 \ifnum #1\expandafter \@firstoftwo
 \else \expandafter \@secondoftwo
 \fi
}%
\providecommand \@ifx [1]{%
 \ifx #1\expandafter \@firstoftwo
 \else \expandafter \@secondoftwo
 \fi
}%
\providecommand \natexlab [1]{#1}%
\providecommand \enquote  [1]{``#1''}%
\providecommand \bibnamefont  [1]{#1}%
\providecommand \bibfnamefont [1]{#1}%
\providecommand \citenamefont [1]{#1}%
\providecommand \href@noop [0]{\@secondoftwo}%
\providecommand \href [0]{\begingroup \@sanitize@url \@href}%
\providecommand \@href[1]{\@@startlink{#1}\@@href}%
\providecommand \@@href[1]{\endgroup#1\@@endlink}%
\providecommand \@sanitize@url [0]{\catcode `\\12\catcode `\$12\catcode
  `\&12\catcode `\#12\catcode `\^12\catcode `\_12\catcode `\%12\relax}%
\providecommand \@@startlink[1]{}%
\providecommand \@@endlink[0]{}%
\providecommand \url  [0]{\begingroup\@sanitize@url \@url }%
\providecommand \@url [1]{\endgroup\@href {#1}{\urlprefix }}%
\providecommand \urlprefix  [0]{URL }%
\providecommand \Eprint [0]{\href }%
\providecommand \doibase [0]{http://dx.doi.org/}%
\providecommand \selectlanguage [0]{\@gobble}%
\providecommand \bibinfo  [0]{\@secondoftwo}%
\providecommand \bibfield  [0]{\@secondoftwo}%
\providecommand \translation [1]{[#1]}%
\providecommand \BibitemOpen [0]{}%
\providecommand \bibitemStop [0]{}%
\providecommand \bibitemNoStop [0]{.\EOS\space}%
\providecommand \EOS [0]{\spacefactor3000\relax}%
\providecommand \BibitemShut  [1]{\csname bibitem#1\endcsname}%
\let\auto@bib@innerbib\@empty
\end{thebibliography}%


\begin{thebibliography}{}
\bibitem{TBG1} R. Bistritzer, A. H. MacDonald, Moir\'{e} bands in twisted double-layer graphene, \emph{Proc. Natl. Acad. Sci. U.S.A.} \textbf{108}, 12233 (2011).\doi{10.1073/pnas.1108174108}.

\bibitem{TBG2} Y. Cao, V. Fatemi, S. Fang, K. Watanabe, T. Taniguchi, E. Kaxiras, and P. Jarillo-Herrero, Unconventional superconductivity in magic-angle graphene superlattices, Nature \emph{(London)} \textbf{556}, 43 (2018).\doi{10.1038/nature26160}.

\bibitem{TBG3} Y. Cao, V. Fatemi, A. Demir, S. Fang, S. L. Tomarken, J. Y. Luo, J. D. Sanchez-Yamagishi, K. Watanabe, T. Taniguchi, E. Kaxiras, R. C. Ashoori, and P. Jarillo-Herrero, Correlated insulator behaviour at half-filling in magic-angle graphene superlattices. \emph{Nature (London)} \textbf{556}, 80 (2018).\doi{10.1038/nature26154}.


\bibitem{TBG4} G.W. Burg, J. Zhu, T. Taniguchi, K. Watanabe, A. H. MacDonald, and E. Tutuc, Correlated Insulating States in Twisted Double Bilayer Graphene, \emph{Phys. Rev. Lett.} \textbf{123}, 197702 (2019).\doi{10.1103/PhysRevLett.123.197702}.

\bibitem{Twistronics} S. Carr, D. Massatt, S. Fang, P. Cazeaux, M. Luskin, and E. Kaxiras, Twistronics: Manipulating the electronic properties of two-dimensional layered structures through their twist angle, \emph{Phys. Rev. B} \textbf{95}, 075420 (2017).\doi{10.1103/PhysRevB.95.075420}.

\bibitem{PL-induced} J.W. Fleischer, M. Segev, N. K. Efremidis, and D. N. Christodoulides, Observation of two-dimensional discrete solitons in optically induced nonlinear photonic lattices, \emph{Nature (London)} \textbf{422}, 147 (2003).\doi{10.1038/nature01452}.


\bibitem{opt1}P. Wang, Y. Zheng, X. Chen, C. Huang, Y. V. Kartashov, L. Torner, V. V. Konotop, and F. Ye, Localization and delocalization of light in photonic moir\'{e}  lattices, \emph{Nature (London)} \textbf{577}, 42-46 (2020).\doi{10.1038/s41586-019-1851-6}.

\bibitem{opt2} Q. Fu, P. Wang, C. Huang, Y. V. Kartashov, L. Torner, V. V. Konotop and F. Ye, Optical soliton formation controlled by angle twisting in photonic moir\'{e}  lattices, \emph{Nat. Photon.} \textbf{14}, 663-668 (2020).\doi{10.1038/s41566-020-0679-9}.

\bibitem{SP16} C. Huang, F. Ye, X. Chen, Y. V. Kartashov, V. V. Konotop, and L. Torner, Localization-delocalization wavepacket transition in Pythagorean aperiodic potentials, \emph{Sci. Rep.} \textbf{6}, 32546 (2016).\doi{10.1038/srep32546 (2016)}.


\bibitem{opt-laser} X.-R. Mao, Z.-K. Shao, H.-Y. Luan, S.-L. Wang, and R.-M. Ma, Magic-angle lasers in nanostructured moir\'{e} superlattice, \emph{Nat. Nanotechnol.} \textbf{16}, 1099 (2021).\doi{10.1038/s41565-021-00956-7}.


\bibitem{Quadratic} Y. V. Kartashov, F. Ye, V. V. Konotop and L. Torner, Multifrequency solitons in commensurate-incommensurate photonic moir\'{e} lattices, \emph{Phys. Rev. Lett.} \textbf{127(16)}, 163902 (2021).\doi{10.1103/PhysRevLett.127.163902}.

%

\bibitem{BEC-moire1}A. Gonz\'{a}lez-Tudela  and J. I. Cirac, Cold atoms in twisted-bilayer optical potentials, \emph{Phys. Rev. A} \textbf{100}, 053604 (2019).\doi{10.1103/PhysRevA.100.053604}.


\bibitem{BEC-moire2} T. Salamon,  A. Celi, R. W. Chhajlany, I.Fr\'{e}rot, Maciej Lewenstein, L. Tarruell, and D. Rakshit, Simulating Twistronics without a Twist, \emph{Phys. Rev. Lett.} \textbf{125}, 030504 (2020).\doi{10.1103/PhysRevLett.125.030504}.


\bibitem{BEC-moire3} X.-W. Luo and C. Zhang, Spin-twisted optical lattices: tunable flat bands and Larkin-Ovchinnikov superfluids, \emph{Phys. Rev. Lett.} \textbf{126}, 103201 (2021).\doi{10.1103/PhysRevLett.126.103201}.

\bibitem{FOP} Z. Chen, X. Liu, and J. Zeng, Electromagnetically induced moir\'{e} optical lattices in a coherent atomic gas \emph{Front. Phys.} \textbf{17}, 42508 (2022).\doi{10.1007/s11467-022-1153-6}.

\bibitem{PCF} Y. S. Kivshar and G. P. Agrawal, \emph{Optical solitons: From fibers to photonic crystals} (Academic, 2003).

\bibitem{PC} J. D. Joannopoulos, S. G. Johnson, J. N. Winn, and R. D. Meade, \emph{Photonic Crystals: Molding the Flow of Light}, 2nd ed. (Princeton University, 2011).

\bibitem{rev-light} I. L. Garanovich, S. Longhi, A. A. Sukhorukova, and Y. S. Kivshar, Light propagation and localization in modulated photonic lattices and waveguides, Phys. Rep. \textbf{518}, 1 (2012).\doi{10.1016/j.physrep.2012.03.005}.

\bibitem{OL-RMP} O. Morsch and M. Oberthaler, Dynamics of Bose-Einstein condensates in optical lattices, \emph{Rev. Mod. Phys.} \textbf{78}, 179 (2006).\doi{10.1103/RevModPhys.78.179}.


\bibitem{NL-RMP} Y. V. Kartashov, B. A. Malomed and L. Torner, Solitons in nonlinear lattices, \emph{Rev. Mod. Phys.} \textbf{83}, 247 (2011).\doi{10.1103/RevModPhys.83.247}.

\bibitem{rev-trapping} Y. V. Kartashov, G. E. Astrakharchik, B. A. Malomed, and L. Torner, Frontiers in multidimensional self-trapping of nonlinear fields and matter, \emph{Nat. Rev. Phys.} \textbf{1}, 185 (2019).\doi{10.1038/s42254-019-0025-7}.



 \bibitem{ABG} I. H. Deutsch, R. J. C. Spreeuw, S. L. Rolston, and W. D. Phillips, Photonic band gaps in optical lattices, \emph{Phys. Rev. A} \textbf{52}, 1394 (1995).\doi{10.1103/PhysRevA.52.1394}.


\bibitem{GS1}  W. Chen and D. L. Mills, Gap solitons and the nonlinear optical response of superlattices, \emph{Phys. Rev. Lett.}  \textbf{58}, 160 (1987).\doi{10.1103/PhysRevLett.58.160}.

\bibitem{GS4} S. John and N. Ak\"{o}zbek, Nonlinear optical solitary waves in a photonic band gap, \emph{Phys. Rev. Lett.}  \textbf{71}, 1168 (1993).\doi{10.1103/PhysRevLett.71.1168}.

\bibitem{GS5}  A. Kozhekin and G. Kurizki,   Self-induced transparency in Bragg reflectors: gap solitons near absorption resonances, \emph{Phys. Rev. Lett.}  \textbf{74}, 5020 (1995).\doi{10.1103/PhysRevLett.74.5020}.


\bibitem{GS6}  A. E. Kozhekin, G. Kurizki, and B. Malomed,   Standing and moving gap solitons in resonantly absorbing gratings, \emph{Phys. Rev. Lett.}  \textbf{81}, 3647 (1998).\doi{10.1103/PhysRevLett.81.3647}.

\bibitem{GS8}  Y. Zhang and B. Wu,   Composition relation between gap solitons and Bloch waves in nonlinear periodic systems, \emph{Phys. Rev. Lett.}  \textbf{102}, 093905 (2009).\doi{10.1103/PhysRevLett.102.093905}.

\bibitem{GS9} E. A. Ostrovskaya and Y. S. Kivshar, Matter-wave gap solitons in atomic band-gap structures, \emph{Phys. Rev. Lett.} \textbf{90}, 160407 (2003).\doi{10.1103/PhysRevLett.90.160407}.


%

\bibitem{BEC-darkGS} L. Zeng and J. Zeng, Gap-type dark localized modes in a Bose-Einstein condensate with optical lattices, \emph{Adv. Photonics} \textbf{1}, 046006 (2019).\doi{10.1117/1.AP.1.4.046004}.
\bibitem{NL-focus} J. Shi and J. Zeng, Self-trapped spatially localized states in combined linear-nonlinear periodic potentials, \emph{Front. Phys.} \textbf{15}, 12602 (2020).\doi{10.1007/s11467-019-0930-3}.
\bibitem{DarkGS-Q} J. Li and J. Zeng, Dark matter-wave gap solitons in dense ultracold atoms trapped by a one-dimensional optical lattice, \emph{Phys. Rev. A} \textbf{103}, 013320 (2021).\doi{10.1103/PhysRevA.103.013320}.
\bibitem{DarkGS-CQ} J. Chen and J. Zeng, Dark matter-wave gap solitons of Bose-Einstein condensates trapped in optical lattices with competing cubic-quintic nonlinearities, \emph{Chaos, Solitons \& Fractals} \textbf{150}, 111149 (2021).\doi{10.1016/j.chaos.2021.111149}.





\bibitem{GS-FBG} B. J. Eggleton, R. E. Slusher, C. M. de Sterke, P. A. Krug and J. E. Sipe, Bragg Grating Solitons, \emph{Phys. Rev. Lett.} \textbf{76}, 1627 (1996).\doi{10.1103/PhysRevLett.76.1627}.
\bibitem{GS-WA} D. Mandelik, R. Morandotti, J. S. Aitchison and Y. Silberberg, Gap Solitons in Waveguide Arrays, \emph{Phys. Rev. Lett.} \textbf{92}, 093904 (2004).\doi{10.1103/PhysRevLett.92.093904}.
 \bibitem{GS-BEC} B. Eiermann, T. Anker, M. Albiez, M. Taglieber, P. Treutlein, K. P. Marzlin and M. K. Oberthaler, Bright Bose-Einstein Gap Solitons of Atoms with Repulsive Interaction. \emph{Phys. Rev. Lett.} \textbf{92}, 230401 (2004).\doi{10.1103/PhysRevLett.92.230401}.



\bibitem{Bender-PT} C. M. Bender and S. Boettcher, Real spectra in non-Hermitian Hamiltonians having $\mathcal{PT}$ symmetry, \emph{Phys. Rev. Lett.}  \textbf{80}, 5243 (1998).\doi{10.1103/PhysRevLett.80.5243}.
\bibitem{Bender-Review} C. M. Bender, Making sense of non-Hermitian Hamiltonians,  \emph{Rep. Prog. Phys.}  \textbf{70}, 947 (2007).\doi{10.1088/0034-4885/70/6/R03}.


\bibitem{NPhoton-PT}  L. Feng, R. El-Ganainy and L. Ge, Non-Hermitian photonics based on parity-time symmetry, \emph{Nat. Photon.} \textbf{11}, 752 (2017).\doi{10.1038/s41566-017-0031-1}.
\bibitem{NSR-PT} H. Zhao and L. Feng, Parity-time symmetric photonics, \emph{Natl. Sci. Rev.}  \textbf{5}, 183 (2018).\doi{10.1093/nsr/nwy011}.
\bibitem{NM-PT}  S. K. \"{O}zdemir , S. Rotter, F. Nori, and L. Yang, Parity-time symmetry and exceptional points in photonics, \emph{Nat. Mater.} \textbf{18}, 783 (2019).\doi{10.1038/s41563-019-0304-9}.
\bibitem{AM-PT}  S. K. Gupta, Y. Zou, X.-Y. Zhu, M.-H. Lu, L.-J. Zhang, X.-P. Liu, and Y.-F. Chen, Parity-time symmetry in non-Hermitian complex optical media, \emph{Adv. Mater.}  \textbf{32}, 1903639 (2020).\doi{10.1002/adma.201903639}.

\bibitem{PT-OL-3level} C. Hang, G. X. Huang, and V. V. Konotop, $\mathcal{PT}$ symmetry with a system of three-level atoms, \emph{Phys. Rev. Lett.} \textbf{110}, 083604 (2013).\doi{10.1103/PhysRevLett.110.083604}.

 \bibitem{PT-atomOL-exp1} Z. Zhang, Y. Zhang, J. Sheng, L. Yang, M.-A. Miri, D. N. Christodoulides, B. He, Y. Zhang, and M. Xiao, Observation of parity-time symmetry in optically induced atomic lattices,  \emph{Phys. Rev. Lett.} \textbf{117}, 123601 (2016).\doi{10.1103/PhysRevLett.117.123601}.


\bibitem{Soliton-PT} Z. H. Musslimani, K. G. Makris, R. El-Ganainy, and D. N. Christodoulides, Optical solitons in $\mathcal{PT}$  periodic potentials, \emph{Phys. Rev. Lett.} \textbf{100}, 030402 (2008).\doi{10.1103/PhysRevLett.100.030402}.

 \bibitem{PRE-PT} J. Zeng and Y. Lan, Two-dimensional solitons in $\mathcal{PT}$ linear lattice potentials,  \emph{Phys. Rev. E}  \textbf{85}, 047601 (2012).\doi{10.1103/PhysRevE.85.047601}.


\bibitem{RMP-PT} V. V. Konotop, J. Yang and D. A. Zezyulin, Nonlinear waves in $\mathcal{PT}$-symmetric systems, \emph{Rev. Mod. Phys.} \textbf{88}, 035002 (2016).\doi{10.1103/RevModPhys.88.035002}.

\bibitem{LPR-PT} S. V. Suchkov, A. A. Sukhorukov, J. Huang, S. V. Dmitriev, C. Lee and Y. S. Kivshar, Nonlinear switching and solitons in $\mathcal{PT}$-symmetric photonic systems, \emph{Laser Photon. Rev.} \textbf{10}, 177 (2016).\doi{10.1002/lpor.201500227}.

\bibitem{Isci} Jiawei Li, Yanpeng Zhang, and Jianhua Zeng, Matter-wave gap solitons and vortices in three-dimensional parity-time-symmetric optical lattices, \emph{iScience.} \textbf{25}, 104026 (2022).\doi{10.1016/j.isci.2022.104026}.


\bibitem{NLS-GP} D. E. Pelinovsky, \emph{Localization in Periodic Potential: From Schr\"{o}dinger Operators to the Gross-Pitaevskii Equation} (Cambridge University Press, Cambridge, 2011).

\bibitem{MSOM} J. Yang, \textit{Nonlinear Waves in Integrable and Nonintegrable Systems}, (SIAM: Philadelphia, 2010).
 \end{thebibliography}
\end{document}